\documentclass[12pt]{iopart}
\usepackage{graphicx}
\usepackage{amssymb}
\usepackage{epsfig}
\usepackage{rotating}
\usepackage{epsfig}
\usepackage{amsfonts, amsbsy, pstricks, pst-node, pst-text}
\usepackage{hyperref}

\def\rmsf#1{{\rm \sf #1}}
\newcommand{\A}{{\rmsf A}}
\newcommand{\pT}{p_{_\perp}}
\newcommand{\xt}{x_{_\perp}}
\newcommand{\meaneps}{{\langle\epsilon\rangle}}
\def\sqrts{{\sqrt{s}}}
\newcommand{\sqrtsnn}{\sqrt{s_{_{\mathrm{NN}}}}}
\def\cO#1{{{\cal{O}}}\left(#1\right)}
\def\eps{\epsilon}
\def\lQCD{\Lambda_{_{\rm QCD}}}
\def\raa{R_{_{\A\A}}}
\def\Ncoll{\langle N_{_{\rm coll}} \rangle}
\def\Npart{\langle N_{_{\rm part}} \rangle}
\def\pp{$p$--$p\ $}
\def\taa{{\langle T_{_{\rmsf{AA}}}\rangle}_{_{\cal C}}}

\begin{document}

\title[Quenching of hadron and photon spectra from RHIC to LHC]{Quenching of hadron and photon spectra in heavy-ion collisions from RHIC to LHC}

\author{Fran\c{c}ois Arleo}

\address{Laboratoire d'Annecy-le-Vieux de Physique Th\'eorique (LAPTH), UMR5108, Universit\'e de Savoie, CNRS, BP 110, 74941 Annecy-le-Vieux cedex, France}
\ead{arleo@lapp.in2p3.fr}
\begin{abstract}
	The generic features of parton energy loss effects on the quenching of single hadron spectra in heavy-ion collisions are discussed, paying attention to the expected differences from RHIC to LHC.
 The need for precise baseline measurements in $p$--Pb collisions at the LHC is also emphasized. Finally I briefly mention the production of prompt photons in heavy-ion collisions as well as some open questions.
	\end{abstract}

\section{Introduction}

The quenching of large-$\pT$ hadrons in heavy-ion collisions 
is undoubtedly one of the most important discoveries obtained so far at RHIC energies ($\sqrtsnn=200$~GeV), suggesting significant parton energy loss in the dense QCD medium produced in those reactions~\cite{dEnterria:2009am}. It was therefore quite natural to wonder whether a similar phenomenon would be observed, either qualitatively or quantitatively, in the first Pb--Pb collisions obtained at LHC at a center-of-mass energy of $\sqrtsnn=2.76$~TeV. Even though \pp data at this energy were then desperately lacking, the first ALICE measurements~\cite{Aamodt:2010jd} soon indicated that the quenching of 
inclusive 
charged hadrons is indeed similar to what has been reported at RHIC, at least in a large $\pT$ domain ($5\lesssim \pT \lesssim 20$~GeV) and within the experimental uncertainties. These first LHC data have later been confirmed by the ATLAS~\footnote{In the case of ATLAS data, only the central-to-peripheral ratio has been measured.}~\cite{Steinberg:2011dj}, ALICE~\cite{aliceraa}, CMS~\cite{Lee:2011cs} preliminary measurements shown at this  conference, which now extend up to $\pT\simeq100$~GeV. The results obtained at the LHC  --~together with the RHIC measurements~-- thus open up hope for a better understanding of energy loss processes through detailed phenomenological studies.

\section{Quenching factor, $\raa$}

The quenching factor is defined as the normalized yield of particles
 in heavy-ion collisions (in a given centrality class ${\cal C}$) over the production cross section measured in \pp collisions,
\begin{equation}\label{eq:defraa}
\raa(\pT, y, {\cal C}) =  \frac{1}{\taa} \times \frac{d^2N^{\A\A}_{_{\cal C}}}{d\pT dy}\ \bigg/ \frac{d^2\sigma^{pp}}{d\pT dy},
\end{equation}
where the thickness function $\taa$ is determined from a Monte Carlo Glauber model. The uncertainty on $\taa$ therefore translates into an overall normalization error
 on $\raa$, which is of the order of 8\% in central Pb--Pb collisions~\cite{Aamodt:2010jd}. Assuming that
the production of \emph{hard} processes, i.e. with a large energy scale $\pT, M \gg \lQCD$, scales with the number of binary nucleon-nucleon collisions ${\Ncoll}_{_{\cal C}}=\taa\times\sigma_{_{\rmsf{NN}}}$, the quenching factor~(\ref{eq:defraa}) becomes $\raa\simeq1$ in the absence of cold and hot medium effects.
 On the contrary the yield of \emph{soft} particles (e.g. light hadron production at small $\pT$) is expected to be proportional to the number of participating nucleons in the collision, $\Npart$, leading to $\raa\simeq \Npart/\Ncoll\simeq A^{-1/3}$ in central A--A collisions.
This conventional assumption, not obvious \emph{a priori}~\footnote{Amusingly, the medium-induced gluon spectrum is proportional to $\Ncoll$ for soft radiation and proportional to $\Npart$ for hard radiation, due to the LPM coherence effect~\cite{Dokshitzer:2004tp}, offering a counter-example.}, remains to be checked experimentally for a variety of ``baseline'' processes insensitive to medium effects. Even though observing binary scaling for one hard process does not prove that this assumption is valid for another, at least this would give some confidence that this is indeed the case.

The quenching factor 
also requires the knowledge of the \pp production cross section at the same collision energy, as is the case at RHIC. At LHC, various interpolation procedures using \pp data 
at $\sqrts=0.9$ and 7~TeV have been performed~\footnote{As a cross-check, Pb--Pb data have
also been compared to the preliminary spectrum measured in the $\sqrts=2.76$~TeV p--p run in 2011.}
assuming a power-law behavior for the $\sqrts$-dependence of the cross section at fixed $\pT$\cite{aliceraa} or fixed $\xt$\cite{Lee:2011cs,collaboration:2011qp}. The latter is probably more appropriate~\cite{Arleo:2010kw} since at large $\pT$ the invariant production cross section is expected to scale like $F(\xt)/\sqrts^n$. Moreover, the $\pT$ interpolation 
is problematic from a statistical point of view since this procedure is limited to the maximal $\pT$-bin available at the \emph{lower} center-of-mass energy, for which phase-space is more restricted. In order to check how these two interpolation prescriptions differ, I compared the yields at $\pT=10$~GeV at $\sqrts=2.76$~TeV using the CMS data within those two procedures. Remarkably, the yields differ by as much as $\sim 25\%$, the $\pT$-interpolation leading to a yield \emph{smaller} than the $\xt$-interpolation procedure. As a consequence, if \pp data indeed follow $\xt$ scaling in the $\sqrts=0.9$--7~TeV range, the measured yield at $\sqrts=2.76$~TeV should be larger than the ALICE ($\pT$-interpolated) yield by $25\%$, thus leading to $\raa$ values \emph{smaller} by this amount~\footnote{The first measurements using the 2.76~TeV \pp data confirm this expectation~\cite{aliceraa}.}. This illustrates well the need for \pp and A--A measurements at the same energy.

\section{Reference processes and baseline measurements}

The abundant production of hard processes in \pp collisions at the LHC offers an outstanding opportunity for QCD studies. Processes with many particles in the final state allow for probing multi-leg processes, while more inclusive observables set tight constraints on parton distribution functions (PDF) and fragmentation functions (FF) but also serve as \emph{reference processes} to which heavy-ion measurements can be compared. Remarkably, the first LHC \pp measurements of prompt photons~\cite{Khachatryan:2010fm,Aad:2010sp} and jets~\cite{atlas:2010wv} proved in excellent agreement with QCD expectations. On the contrary, discrepancy between data and NLO calculations were reported in the hadron production channel~\cite{Reygers:2011ny}. The origin of this discrepancy is not clear yet, although it is worth stressing that FF are rather poorly constrained, especially in the gluon sector and at large momentum fraction $z$.
Higher-twist processes might also contribute substantially to hadron production at not too large $\pT$~\cite{Arleo:2009ch}. It might be tempting to assume that the present uncertainties in the absolute \pp cross sections are not an issue in heavy-ion collisions since these most likely cancel out in Eq.~(\ref{eq:defraa}). As we shall see 
later, 
this is not necessarily the case, since the medium effects which affect $\raa$ actually depend on the dynamics of the hard process itself. It is thus important to keep in mind such uncertainties when discussing the various phenomenological studies and their assumptions.

In heavy-ion collisions, measurements are made more delicate because of the larger underlying event. Thanks to 
recent 
jet reconstruction techniques in high-multiplicity environments~\cite{Salam:2009jx}, it has become possible to measure the 
$\raa$
of jets, at RHIC and now at LHC~\cite{Caines:2011xp}. These results nicely supplement 
hadron production data, as they allow for tracing more efficiently the dynamics of parton energy loss and induced gluon radiation in a medium~\footnote{For a discussion on jets 
 in heavy-ion collisions, see~\cite{Zapp:2011kj}.}.
Although energy loss processes are expected to play an important role in the quenching of hadrons and jets, other 
effects could actually affect $\raa$: nuclear PDF (nPDF) and saturation effects, multiple scattering in nuclei and recombination processes in QGP,
to cite a few examples. Hopefully, most of these processes die out at transverse momenta at which the sole effects of parton energy loss remain visible. It is e.g. the case of nPDF effects which become small at large $\pT$~\footnote{Saturation effects should disappear above $\pT\gtrsim Q_{_s}^2/\lQCD\simeq10$--20~GeV where $Q_{_s}\simeq2$~GeV at LHC.}, below 20\% above $\pT\gtrsim 10$~GeV according to EPS09~\cite{Eskola:2009uj}.
In order to disentangle these mechanisms, it is essential to compare systematically hadron/jet 
data to baseline measurements:
\begin{itemize}
\item \emph{Baseline experiments}, typically deep inelastic scattering in nuclei ($e$--A) or proton--nucleus ($p$--A) collisions. These reactions allow for probing initial-state effects, such as nPDF effects or parton propagation in a cold medium;
\item \emph{Baseline observables} in heavy-ion collisions, such as prompt photon, Drell-Yan pair ($\gamma^\star$) and $W^\pm/Z^0$ production, which are expected to be insensitive to medium effects. Yet, they are essential since they offer a ``normalization'' of $\raa$ in central heavy-ion collisions and serve as tests of the binary collision scaling assumption.
\end{itemize}
The need for baseline measurements is best exemplified at RHIC, where 
$\raa$
of hadrons in $d$--Au collisions and that of photons in Au--Au collisions strongly suggest that the hadron quenching at RHIC is a genuine phenomenon (i.e. not a normalization artifact) due to a final-state effect. At LHC, the situation is extremely encouraging regarding the production of electroweak observables, as shown by the first $Z^0$ and isolated  photon measurements by CMS~\cite{collaboration:2011fa}.
As discussed in~\cite{Salgado:2011wc}, a dedicated $p$--Pb run at the LHC is needed and hopefully will be carried out before shutdown in 2013.

\section{Parton energy loss effects on $\raa$}

Before discussing more specific models, I discuss in this section qualitative --~but hopefully  model-independent~-- features of parton energy loss on $\raa$. Let us assume for simplicity that the 
 rate of $y=0$ hadron production~\footnote{Strictly speaking Eq.~(\ref{eq:model}) would be more appropriate for jets than for single-inclusive hadron production, but this is unimportant in this discussion.} in A--A collisions is given by
\begin{equation}\label{eq:model}
\frac{1}{\Ncoll}\ N^h_{AA}(\pT)=\int d\eps\ {\cal P}(\eps)\ N^h_{pp}(\pT+\eps)
\end{equation}
where $N^h_{pp}$ is the corresponding rate in \pp collisions and ${\cal P}(\epsilon)$ the energy loss probability distribution.
 Taylor expanding $N^h_{pp}$ in $\epsilon/\pT$, the quenching factor reads to first order
\begin{equation}\label{eq:raamodel}
\raa^h(\pT)\simeq 1 - (1-p_0)\ \frac{\meaneps(\pT)}{\pT}\ n^h(\xt), \qquad n^h(\xt) \equiv \bigg|\frac{d \ln N^h_{pp}(\pT)}{d \ln \pT}\bigg|,
\end{equation}
where $p_0$ is the probability for no energy loss.
 The index $n$ is the power-law exponent of the \pp rate, which is a scaling function of $\xt$ (up to QCD scaling violations), almost independent of $\xt$ at small $\xt$ and getting larger as $\xt$ goes to 1
  (see Fig.~3 of~\cite{Horowitz:2011gd}). Its value  depends on the slope of the FF into hadrons, consequently (i) the uncertainties on the FF at large $z$ (thus on $n$) will translate into an uncertainty on $\raa$, (ii) a stronger suppression in the baryon channel, e.g. $\raa^p<\raa^\pi$, is in principle expected since
the spectra of baryons are softer than that of mesons. 
\setcounter{footnote}{0}
Apart from the dependence of $n$ on $\xt$,
 $\raa$ depends on the \emph{fractional} energy loss, $\meaneps(\pT)/\pT$. The important lever arm in $\pT$ at the LHC
 therefore allows for probing the energy dependence of $\meaneps$. Let us illustrate this by assuming that $\meaneps(\pT)\propto \pT^\alpha$ and fixing {$\raa(\pT=20$~GeV$)=0.4$}, as suggested by the first LHC measurements. From Eq.~(\ref{eq:raamodel}) and neglecting the  $\pT$ dependence of $n$, the ``predicted'' $\raa$ at $\pT=100$~GeV would range from $\raa\simeq 0.9$ to $\raa=0.4$ when the parameter $\alpha$ is varied from 0 to 1~\footnote{In various QCD approaches, $\meaneps(\pT)\sim\log(\pT)$ leading to an effective exponent $\alpha\simeq 0.3$.}.  Provided $\alpha<1$, note that $\raa\to1$ 
 at asymptotic energies $\pT\to\infty$~\footnote{Here I actually mean $\pT\to\infty$ with $\pT/\sqrts$ fixed, i.e. the edge of phase-space is not reached.}. It would therefore be interesting to measure the values of $\pT^{\rmsf{max}}$ for which  energy loss effects are  negligible, i.e. $\raa(\pT^{\rmsf{max}})\simeq1$, for various centralities in order to constrain the amount of energy lost in the medium.

The large phase-space available at the LHC is also very satisfactory from a theoretical point of view as it allows for a better separation of the different energy scales involved in the process. The ``ideal playground'' for energy loss studies could be summarized as 
$
\lQCD \ll \meaneps \ll \pT \ll \sqrts/2
$.
The first inequality is likely to be fulfilled if the medium is dense or thick enough; phenomenological estimates on the transport coefficient $\hat{q}$ extracted at RHIC 
indicate that it is actually the case. More problematic is the fact that the typical energy loss is of the order of the energy of the hard parton, $\meaneps\simeq\pT$, leading on average to a full stopping of the propagating parton and therefore to strong surface bias effects~\cite{Eskola:2004cr}.
This is a severe limitation in particular at RHIC for which transverse momenta are presently restricted to $\pT\lesssim20$~GeV. The rightmost inequality is perhaps less crucial, yet it is likely to be responsible for the observed flatness of $\raa$ as a function of $\pT$ at RHIC energies: as $\xt\to1$, the index $n(\xt)$ gets larger and compensates the decrease of $\meaneps/\pT$ at large $\pT$. At the LHC, this region is not probed unless hadrons are measured at forward rapidities, for which $\xt\times\exp(y)=\cO{1}$.

The hadron quenching in heavy-ion collisions has now been observed at two different colliding energies, offering a promising possibility to understand further the underlying phenomena from a systematic comparison of RHIC and LHC. One natural possibility is to compare $\raa$ at fixed $\pT$ and in the same centrality class, as the propagating partons would have typically the same energy and would cover the same medium length $L$, despite the larger $\hat{q}$ at LHC~\footnote{$\hat{q}$ is expected to be proportional to the particle density, and $dN/dy\big|_{\rmsf{LHC}}\simeq 2.4\  dN/dy\big|_{\rmsf{RHIC}}$~\cite{Aamodt:2010jd}.}. Similarly one could also compare $\raa$ at fixed $\pT$ and particle density, leading to the same $\hat{q}$ but different $L$. Finally, comparing data at fixed $\xt$ would be interesting since the partonic slopes and nPDF/saturation effects would be identical, yet parton energies would be rather different.

\section{Phenomenology}

\begin{figure}[t]
  \begin{minipage}[t]{8.cm}
    \begin{center}
      \includegraphics[width=8.cm,height=6.8cm]{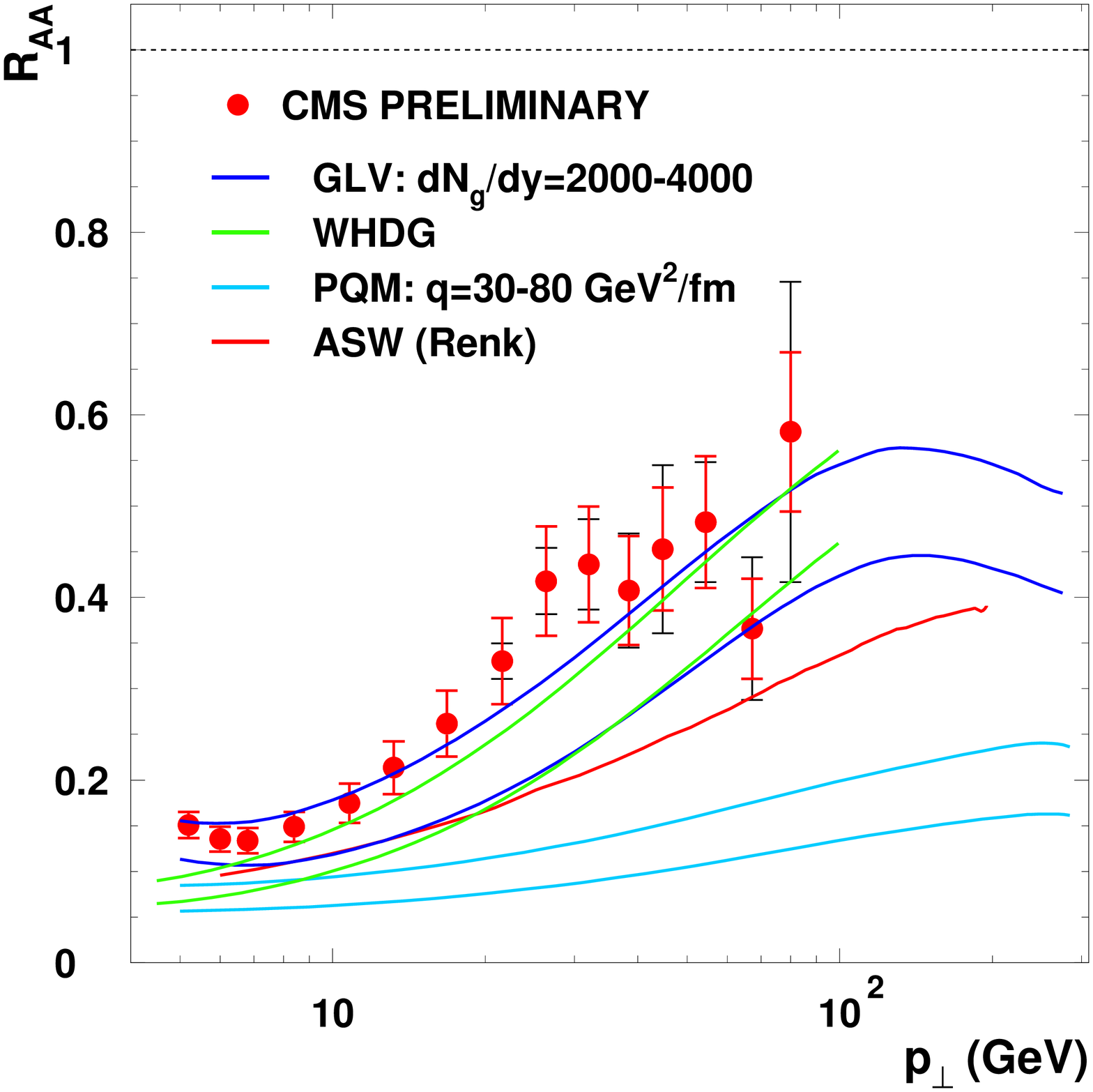}
    \end{center}
    \caption{$\raa^h$ measured by CMS.}
     \label{fig:raadata}
  \end{minipage}
\hfill
  \begin{minipage}[t]{8.cm}
    \begin{center}
      \includegraphics[width=8.cm,height=6.8cm]{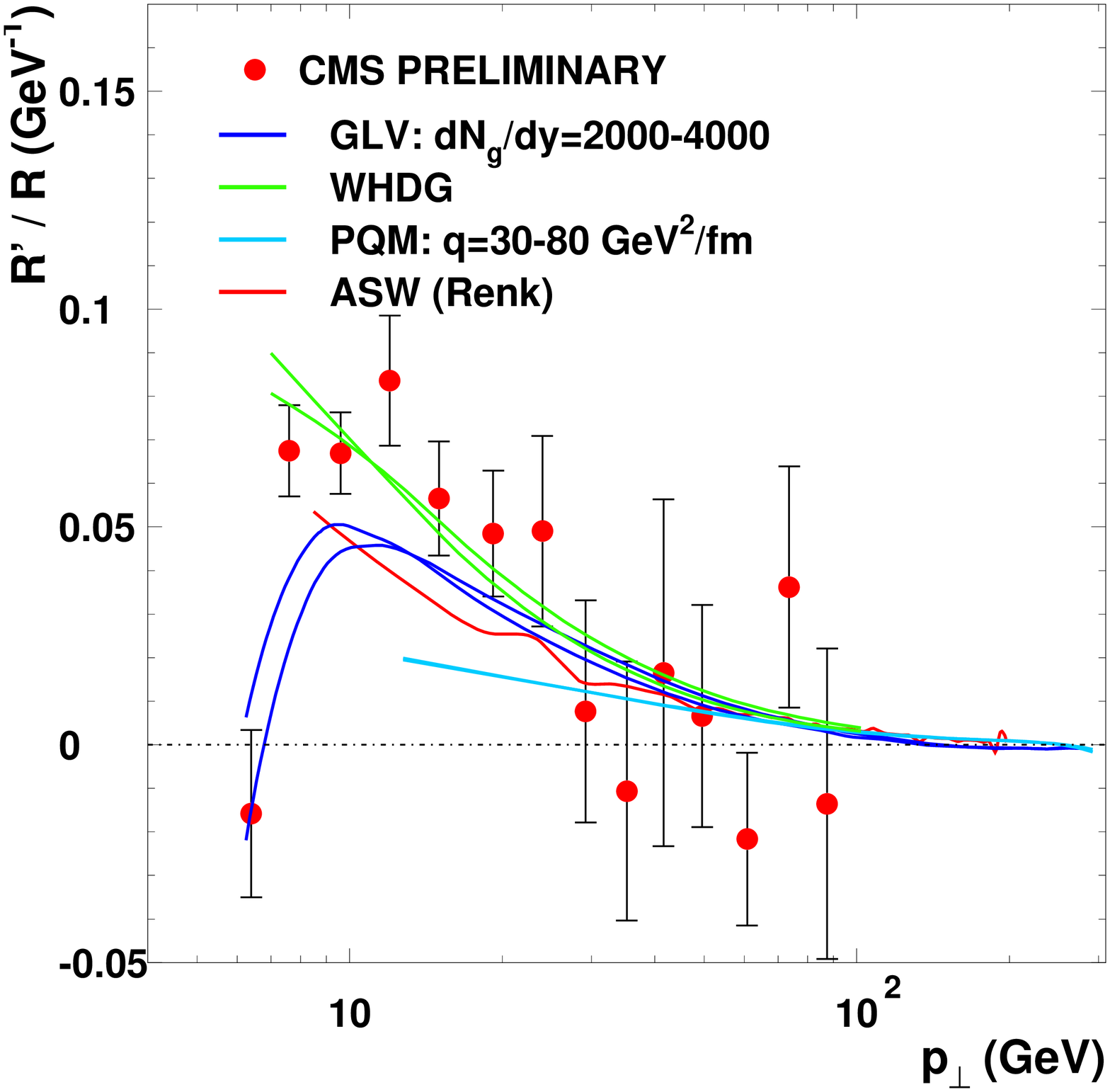}
    \end{center}
    \caption{${R^\prime_{_{\rmsf{log}}}}$ extracted from CMS data.}
 \label{fig:lograadata}
  \end{minipage}
\end{figure}

There is a plethora of phenomenological studies on $\raa$ at RHIC/LHC, mostly based on parton energy loss.
The strategy usually followed can be summarized as follows:
\begin{enumerate}
\item[(i)] A general energy loss framework is chosen. To be schematic, one usually distinguishes the approaches in which propagating partons experience soft multiple scattering (BDMPS-ASW \& AMY) to the ones in which only few hard scatterings take place (DGLV \& higher-twist (HT)). The BDMPS-ASW and DGLV set-ups assume static scattering centers, while AMY applies to thermal media. In the HT approach the amount of energy loss is related to the strength of quark-gluon correlations in the medium. Additional work is then needed to treat \emph{multiple} gluon emission, assuming independent (Poisson) gluon emission (BDMPS-ASW \& DGLV),  solving rate equations (AMY) or virtuality-ordered, DGLAP-like, evolution equations (HT).  Comparing these frameworks quantitatively has been the goal of the TECHQM collaboration over the last few years~\footnote{\url{https://wiki.bnl.gov/TECHQM}}. For more details, see~\cite{Armesto:2011ht};
\item[(ii)] Apart from radiative energy loss, other medium effects are sometimes implemented, e.g. collisional energy loss --~which can be important at not too large $\pT$ and for heavy-quarks~-- or nPDF effects;
\item[(iii)] Additional assumptions are made regarding the produced medium and its hydrodynamical evolution. For a long time the medium expansion was described according to the simple Bjorken model, i.e. a pure longitudinal expansion. More recently, energy loss models have been implemented into more sophisticated 2D (w/o viscous corrections) or 3D hydro evolution. In most studies, the amount of energy loss is fixed at RHIC and scaled up at the LHC according to the particle density.
\end{enumerate}
In Fig.~\ref{fig:raadata} are compared the preliminary CMS data~\cite{Lee:2011cs} with various model predictions, either based on BDMPS-ASW (PQM~\cite{Dainese:2004te} \& Renk~\cite{Renk:2011gj}) or DGLV (GLV~\cite{Vitev:2002pf}\& WHDG~\cite{Horowitz:2011gd}) frameworks~\footnote{Other predictions based on the HT framework were made in~\cite{Chen:2011vtMajumder:2011uk}.}. The bands indicate the uncertainties coming from the initial particle density assumed in each model. Apart from PQM,
 the various models predict a rising $\raa$ with $\pT$ in qualitative agreement with data.  Even though GLV \& WHDG calculations best reproduce the magnitude of the data, no premature conclusion should be drawn at this stage regarding the other two predictions.
As mentioned previously, the $\pT$ \emph{slope} of $\raa$ at the LHC offers an interesting possibility to probe energy loss processes. On the contrary, the \emph{magnitude} of $\raa$ is useful in order to calibrate the strength of energy loss in the medium. For this reason, it might be worth considering also the logarithmic derivative of $\raa$, 
\begin{equation}\label{eq:raaprime}
R^\prime_{_{\rmsf{log}}} \equiv (\log \raa)^\prime = \frac{d \ln \raa}{d\pT},
\end{equation}
which quantifies the slope of $\raa$ with $\pT$ and which is free of normalization uncertainties.
 This is illustrated in Fig.~\ref{fig:lograadata}, where $R^\prime_{_{\rmsf{log}}}$ has been extracted from both data and theory~\footnote{I thank E. Wenger and A.S. Yoon for providing me with the CMS preliminary data in Fig.~\ref{fig:raadata}.}. 
Remarkably, the theoretical uncertainties shrink since the bands in Fig.~\ref{fig:raadata}  essentially reflect a normalization uncertainty on $\raa$. This variable is thus very much independent on the \emph{amount} of energy loss but tests the dynamics of energy loss models. At large $\pT\gtrsim30$~GeV, data and models (including PQM) agree. Note also that $R^\prime_{_{\rmsf{log}}}$ is consistent with zero, indicating a flat behavior of $\raa$ in this $\pT$ domain. At lower $\pT$, the model which describes best the data is WHDG, while Renk, GLV and PQM tend to underpredict the slope of $\raa$. This discussion is not meant to be comprehensive, but is rather an illustration of the potential usefulness of quantifying the slope of $\raa$ through Eq.~(\ref{eq:raaprime}). Since $R^\prime_{_{\rmsf{log}}}$ is dominated by statistical uncertainties (unlike $\raa$ at small $\pT$), $R^\prime_{_{\rmsf{log}}}$ is expected to become more precise as more data are collected.

\section{Photons}

Prompt photons have long been thought to be \emph{the} ideal baseline measurement in heavy-ion collisions to which hadron production $\raa$ should be contrasted. They can be used to probe nuclear PDF~\cite{Arleo:2007js}
at not too large $\pT$ while at larger $\pT$ (say, $\pT\gtrsim50$~GeV) they offer a good test of binary collision scaling. However, it soon became clear that various hot medium effects could actually affect photon production in nuclear collisions~\cite{Gale:2009gc}:
\begin{itemize}
\item Jet-to-photon conversion, in which a produced high-energy parton converts all of its momentum to a photon by a rescattering in the medium;
\item Photon bremsstrahlung induced by parton multiple scattering in the medium;
\item Quenching of the fragmentation component, similar to the quenching of hadrons due to energy loss effects.
\end{itemize}
\setcounter{footnote}{0}
Both jet conversion and photon bremsstrahlung should \emph{enhance} photon production leading to $\raa>1$~\footnote{In the photon channel and at large $\xt$, $\raa$ should not be compared to unity since the neutrons in nuclei are less efficient to produce photons.}
 and to negative second-Fourier component, $v_{_2}<0$, in non-central collisions. On the contrary, one would expect $\raa<1$ and $v_{_2}>0$ if fragmentation photons are quenched. Also note that the last two effects would not show up when triggering on \emph{isolated} photons, i.e. with little hadronic background in their vicinity.
On the experimental side, PHENIX has reported long ago on preliminary measurements of inclusive photon production 
 Au--Au collisions~\cite{Sakaguchi:2008ec}. These data are consistent with no medium effects within the uncertainties, yet they cannot be excluded. Hopefully the current situation will soon be clarified when final data come out. STAR also reported on photon anisotropies $v_{_2}\gtrsim0$ at large $\pT$~\cite{Hamed:2010ew}.  At the LHC, CMS reported in this conference the first measurement of isolated photon production in heavy-ion collisions~\cite{Lee:2011cs}, which is a remarkable achievement. The data are also consistent with $\raa=1$ although  the present uncertainties are naturally rather large at this stage. 

\section{Open questions}

Instead of a conclusion, let me mention some open questions related to ``jet quenching'' studies and energy loss processes. Regarding the theoretical frameworks themselves, efforts have been made to improve the current approximations, in particular including higher-order corrections and going beyond the soft and collinear limit in the calculation of medium-induced gluon spectrum~\cite{Vitev:2011gs,dEramo} (see also~\cite{Zapp:2011kj} and references therein).
On a more phenomenological side, it would be important to clarify the role of collisional (in contrast to radiative) energy loss effects, in light of the recent LHC data. One surprise of the RHIC data was the absence of proton quenching at rather small $\pT$, whereas in pure energy loss scenarios one would expect a similar (if not stronger) quenching as that of mesons. It would be particularly interesting to observe the quenching of hadrons according to their FF slopes, a phenomenon observed e.g. in DIS on nuclei~\cite{Accardi:2009qv}. The heavy-quark sector should also allow for a critical test of energy loss processes and the possible dead cone hierarchy, $\eps_q>\eps_c>\eps_b$~\cite{Dokshitzer:2001zm}, which seems \emph{a priori} challenged by the ALICE $D$-meson $\raa$~\cite{aliceraa} and by theoretical arguments~\cite{Aurenche:2009dj}. The double ratio $\raa^D/\raa^\pi$ vs. $\pT$ should e.g. set stringent constraints on energy loss models. Another challenge will be disentangle energy loss from possible saturation effects. The fact that the hadron quenching at mid-rapidity is similar at LHC than at RHIC, despite $\xt$ values smaller by an order of magnitude, questions the strength of such saturation effects. Hopefully future $p$--Pb data at LHC may soon clarify this question.

Without a doubt, the large-$\pT$ hadron and photon measurements at the LHC presented at this conference --~as well as that of RHIC with unprecedented luminosities~-- are the first fruits of what will be another exciting decade of tomographic studies.

\section*{References}

\providecommand{\url}[1]{\texttt{#1}}
\providecommand{\urlprefix}{http://arXiv.org/abs/}
\providecommand{\eprint}[2][]{\href{\urlprefix#2}{\url{#2}}}

\end{document}